\documentclass[12pt]{article}
\usepackage{caption}
\usepackage{subcaption}
\usepackage{hstproposaltemplate_cy34}
\usepackage{ragged2e}
\usepackage{amsmath}

\usepackage{graphicx}
\usepackage{xcolor}
\usepackage{hyperref}
\usepackage{natbib}
\usepackage{hstcite}
\usepackage{txfonts}
\usepackage{fancyhdr}
\usepackage{authblk}
\usepackage{orcidlink}
\usepackage[left=1.5cm, right=1.5cm, top=2cm, bottom=2cm]{geometry}
\hypersetup{%
  citecolor=blue
}
\hypersetup{
  pdfborder={0 0 0},
  linkbordercolor=[rgb]{0.5,0.57,0.72},
  linkcolor=[rgb]{0.0,0.0,0.5},
  urlcolor=[rgb]{0.0,0.0,0.5},
  colorlinks=true,
}
\newcommand{\Ion}[2]{#1\,\textsc{#2}}

\begin{document}

\centering\section*{\large White dwarf planetary systems in the ultraviolet}

\textbf{Principal author: Jamie Williams$^{\orcidlink{0009-0007-8709-9689}}$\footnote{jamietwilliams.astro@gmail.com} (University of Warwick)}

\vspace{2mm}
\noindent\textbf{Co-authors: Amy Bonsor$^{\orcidlink{0000-0002-8070-1901}}$ (University of Cambridge), Boris G{\"a}nsicke$^{\orcidlink{0000-0002-2761-3005}}$ (University of Warwick), Joseph Guidry$^{\orcidlink{0000-0001-9632-7347}}$ (Boston University), JJ Hermes$^{\orcidlink{0000-0001-5941-2286}}$ (Boston University), Lou Baya Ould Rouis$^{\orcidlink{0009-0002-6065-3292}}$ (Boston University), Laura Rogers$^{\orcidlink{0000-0002-3553-9474}}$ (NOIRLab), Pier-Emmanuel Tremblay$^{\orcidlink{0000-0001-9873-0121}}$ (University of Warwick), Snehalata Sahu$^{\orcidlink{0000-0002-0801-8745}}$ (University of Warwick), Andrew Swan$^{\orcidlink{0000-0001-6515-9854}}$ (University of Warwick), David Wilson$^{\orcidlink{0000-0001-9667-9449}}$ (University of Colorado), Siyi Xu$^{\orcidlink{0000-0002-8808-4282}}$ (NOIRLab)}

\RaggedRight
\vspace{-2mm}
\section*{Abstract}
\vspace{-2mm}

Almost every known planet host will evolve into a white dwarf, and the surviving planetary material will continue to orbit this stellar remnant. Asteroids perturbed onto star-grazing orbits will become disrupted, forming an accretion disk which causes ``enrichment'' of the otherwise pure hydrogen or helium atmosphere. Measurements of these photospheric abundances give detailed insights into the interior compositions of exo-planetesimals with an accuracy not possible for intact exoplanets around main sequence stars. This method has revealed the diversity of rocky material in our solar neighborhood, including primitive, chondritic planetesimals, fragments of planetary cores, and even analogues of Kuiper belt objects. The planetesimal abundances can be used as an input to interior structure models. The far-ultraviolet is a key wavelength range for this field because it contains strong transitions for almost every element of interest, many of which are undetectable using ground-based optical spectroscopy. Without the FUV, we will no longer have access to the C, N, P, S content of exoplanetary bodies and thus will no longer be able to probe how volatiles interact with refractories, which is crucial to understanding planet formation$-$and even the origin of life. The medium resolution and high sensitivity of COS on \textit{HST} has been indispensable in determining the compositions of dozens of exo-planetesimals. However, the only two medium resolution far-ultraviolet-capable spectrographs are currently onboard \textit{HST}, with no plans for replacements until the 2040s. An extension to the \textit{HST} mission is critical for the field of white dwarf planetary systems, because the loss of FUV capability would leave us blind to volatiles. Boosting the orbit of \textit{HST} would allow us to measure volatile abundances, determine the rocky planetary occurrence rate, investigate differentiation, and probe for photospheric abundance variability. 

\vspace{-2mm}
\section*{Evolved planetary systems}
\vspace{-2mm}

The transition into a white dwarf is the inevitable conclusion for all isolated main sequence stars with masses $\lesssim8-10\,$M$_\odot$. This transition results in the engulfment of the inner planetary system during the red giant phase. In the Solar System, all planets out to Earth will be destroyed. However, planetary material beyond the maximum radial extent of the star will survive and orbit the white dwarf remnant \citep{Veras2016Review}. The danger does not end however, because surviving planets can perturb material towards the white dwarf \citep{DebesSigurdsson2002,VerasGaensicke2015}. Planetary material which enters the white dwarf's Roche radius is disrupted, forming a debris disk which is gradually accreted. This debris disk is observable as an infrared excess at $\approx1-4\,\%$ of white dwarfs \citep{Jura2003,TWilson2019}, gaseous emission at $\approx0.07\,\%$ of white dwarfs \citep{Gaensicke2006_disc,Manser2020} (and gaseous absorption for favorable inclinations; \citealt{Xu2016}), and transiting debris at $\approx0.3\,\%$ of white dwarfs \citep{Vanderburg2015,Robert2024}. However, the most common signature of planetary accretion is metal pollution of the pure hydrogen or helium atmosphere, observed at $\approx40\,\%$ of warm white dwarfs \citep{Koester2014_DAZ,OuldRouis2024}. Measurements of white dwarf photospheric abundances reveal the composition of the accreted material. This white paper describes scientific goals which can only be realized by ultraviolet spectroscopy using \textit{HST}.

\begin{figure}[t]
    \centering
    \includegraphics[width=\linewidth]{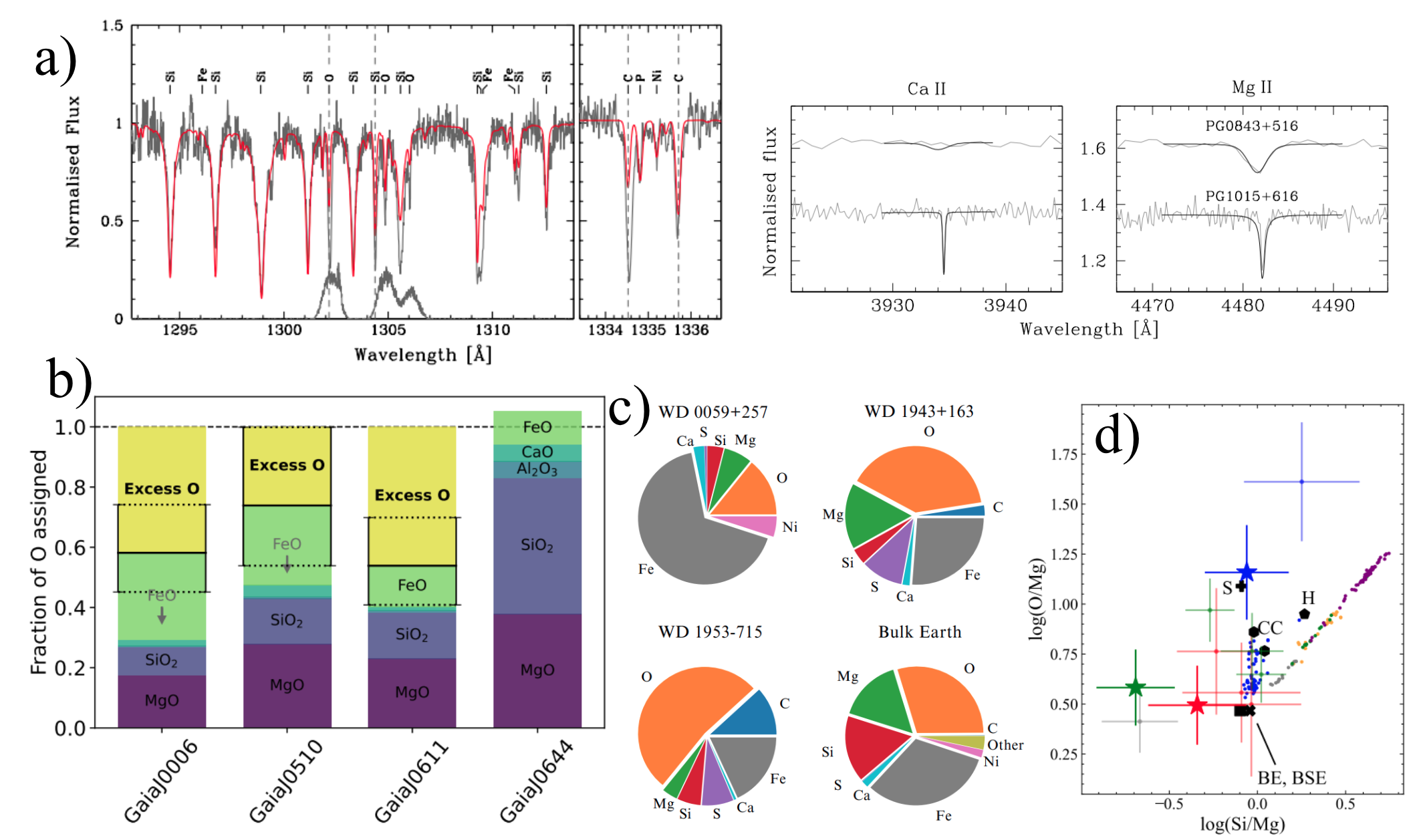}
    \caption{Examples of what is possible with \textit{HST} FUV spectroscopy. Panel a) shows COS spectroscopy of a heavily enriched white dwarf with a myriad of metal lines in the FUV but only Mg and Ca lines in the optical \citep{Gaensicke2012}. Panel b) shows an O budget calculation, finding that three out of the four white dwarfs have excess O indicative of a wet body \citep{Rogers2024b}. Panel c) shows the mass fraction of three white dwarfs accreting a diversity of objects relative to the bulk Earth \citep{Williams2025}. Panel d) shows the number abundance ratios of white dwarfs (points with error bars) compared to Solar System meteorites (small points) \citep{Williams2025}.}
    \label{fig:uvexamples}
\end{figure}

A great diversity of rocky material has been discovered accreting onto white dwarfs \citep{Swan2019}. The far-ultraviolet (FUV) capabilities of \textit{HST} have been instrumental in developing this field, with discoveries including water-rich chondritic asteroids \citep{Hoskin2020}, volatile-rich Kuiper Belt analogues \citep{Xu2017,Sahu2025}, insights into core-rich \citep{Wilson2015,Williams2025} and mantle-rich objects \citep{MelisDufour2017}, and even simultaneous accretion of a rocky body and an icy body \citep{Johnson2022}. The resolution and wavelength coverage of STIS and COS (the most sensitive UV spectrograph ever flown) are vital to this work, as a stellar spectrum with just a few weak features in the optical may be awash with lines from dozens of metals in the FUV (see Fig.\,\ref{fig:uvexamples}).

The commissioning of additional lifetime positions to reduce the effect of gain sag will extend the use of COS into the 2040s, keeping this field alive until the launch of the Habitable Worlds Observatory (HWO). Limitations on count rate on areas of the detector mean that the brightest white dwarfs cannot be observed with COS, but STIS offers an alternative, with the only additional requirement of longer exposure times.  Reduced gyro mode has a negligible effect on the typical spectroscopic observations of this field. These techniques to extend the lifetime of \textit{HST} incur only a small scientific cost to this field, whereas the loss of the observatory would be catastrophic.

\vspace{-2mm}
\section*{Population studies}
\vspace{-2mm}

Unlike transit or radial velocity surveys which are expensive uses of telescope time, a planetary system around a white dwarf is revealed by the presence of metals in its atmosphere, often detectable in just one or two \textit{HST} orbits. The frequency of planet-hosting stars has been constrained by detecting pollution in white dwarfs, with the most sensitive tracer being Si lines in the FUV \citep{Koester2014_DAZ}. The most recent analysis found that $44\pm6\,\%$ of white dwarfs with temperatures $13\,000<T_\mathrm{eff}<30\,000\,$K are actively accreting \citep{OuldRouis2024}. Typical planet detection techniques such as transit or radial velocity surveys are less sensitive around massive stars, but white dwarfs can explore the planetary systems of AFG progenitors.

Of the $\approx1220$ white dwarfs within 100\,pc with $13\,000<T_\mathrm{eff}<30\,000\,$K \citep{GentileFusilloGaia2021} that are ideal targets for FUV observations, only $40\,\%$ have archival FUV spectra obtained by STIS or COS. A volume-limited 100\,pc warm white dwarf FUV sample is ambitious, but obtainable with current \textit{HST} instruments with a mission extension. Not only can we measure an unbiased occurrence rate of warm planetary accreting white dwarfs$-$which is only possible with \textit{HST}$-$but additional heavily enriched white dwarfs will allow us to explore planetary compositions and volatile budgets within the solar neighborhood. White dwarfs can investigate planetary occurrence rate and composition across both stellar mass and age, and the latter can be used as a proxy for metallicity because older stars are generally metal-poor \citep[e.g.][]{Prochaska2000,Bashi2022}.

With a larger sample of heavily enriched white dwarfs with medium resolution FUV spectra, not only can we find the planetary system occurrence rate, but we can put constraints on the composition of different types of planetesimals. Core- or mantle-rich fragments reveal the frequency of differentiation \citep{Bonsor2020}, which can be used to put constraints on the abundance of short-lived radioactive isotopes such as $^{26}$Al in the solar neighborhood \citep{Curry2022}. Primitive, volatile-rich comets likely brought key ingredients to the emergence of life on Earth \citep[e.g.][]{McDonald2025}, and white dwarfs can be used to measure the composition of exo-comets \citep{Xu2017}. Main sequence companions to enriched white dwarfs can be used as a proxy for the initial abundances from which the planet formed \citep{Bonsor2021,Aguilera-Gomez2025}, therefore allowing us to investigate the effect planetary formation and evolution has on planetary composition. A larger population of these planetary accreting white dwarfs observed with \textit{HST} will move the field from interpreting single, interesting objects to a statistical analysis of planetesimal compositions.

More FUV spectra of white dwarfs are necessary to investigate the discrepancy between optical and ultraviolet abundances \citep{Gaensicke2012,Jura2012,Xu2019} as well as discrepancies between ionization states in the ultraviolet \citep{Williams2025}. Solutions to these discrepancies such as varying line formation as a function of depth \citep{Rogers2024a}, physics approximations in white dwarf models \citep{Buchan2025}, or uncertain atomic data \citep{Gaensicke2012} require more white dwarfs with detected Si in the optical and UV, with the UV preferably containing multiple ionization states. By searching for any relationship between larger discrepancies and white dwarf parameters, this will help disentangle tensions between measurements using optical or UV spectra.

\begin{figure}[t]
    \centering
    \includegraphics[width=0.49\linewidth]{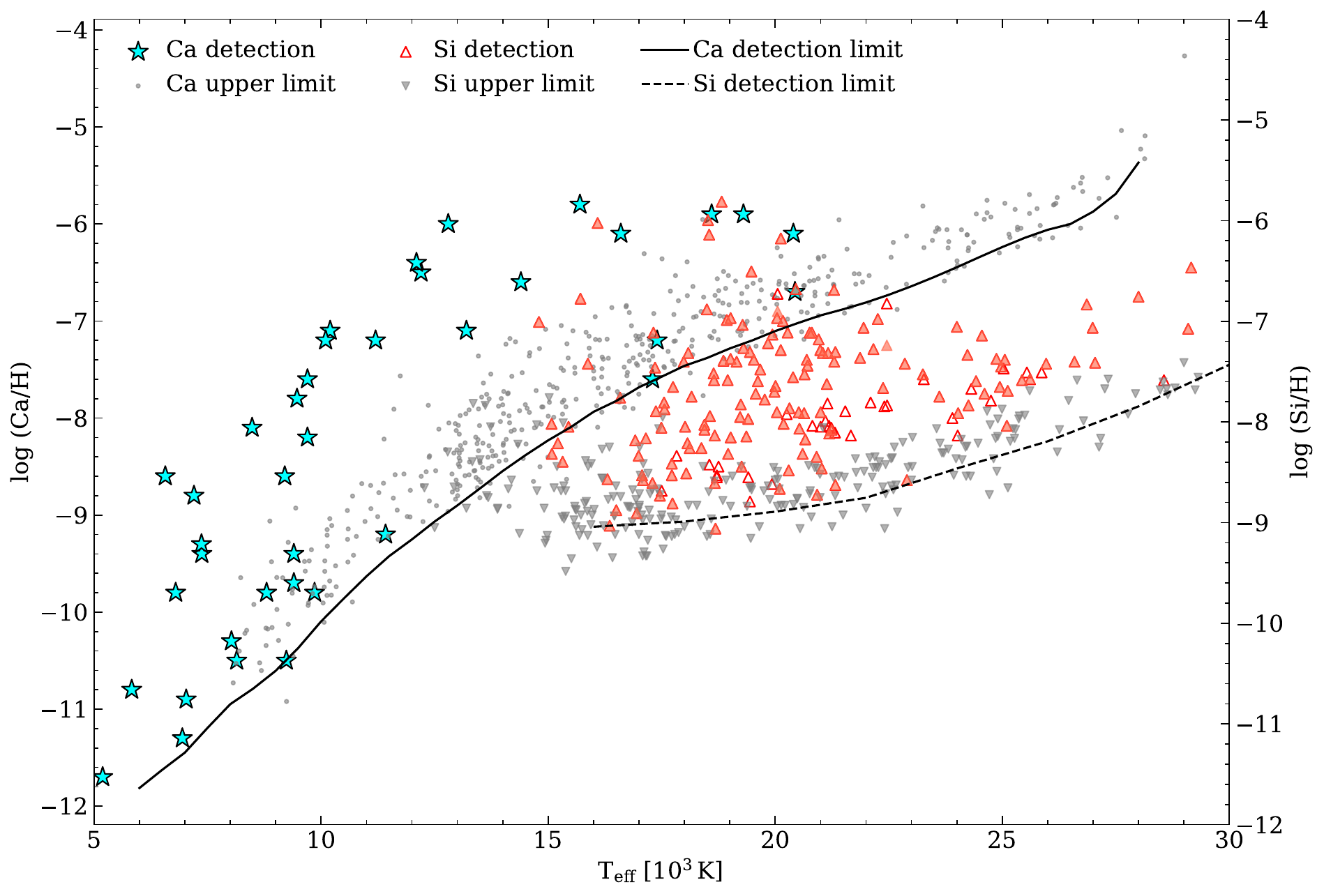}
    \includegraphics[width=0.49\linewidth]{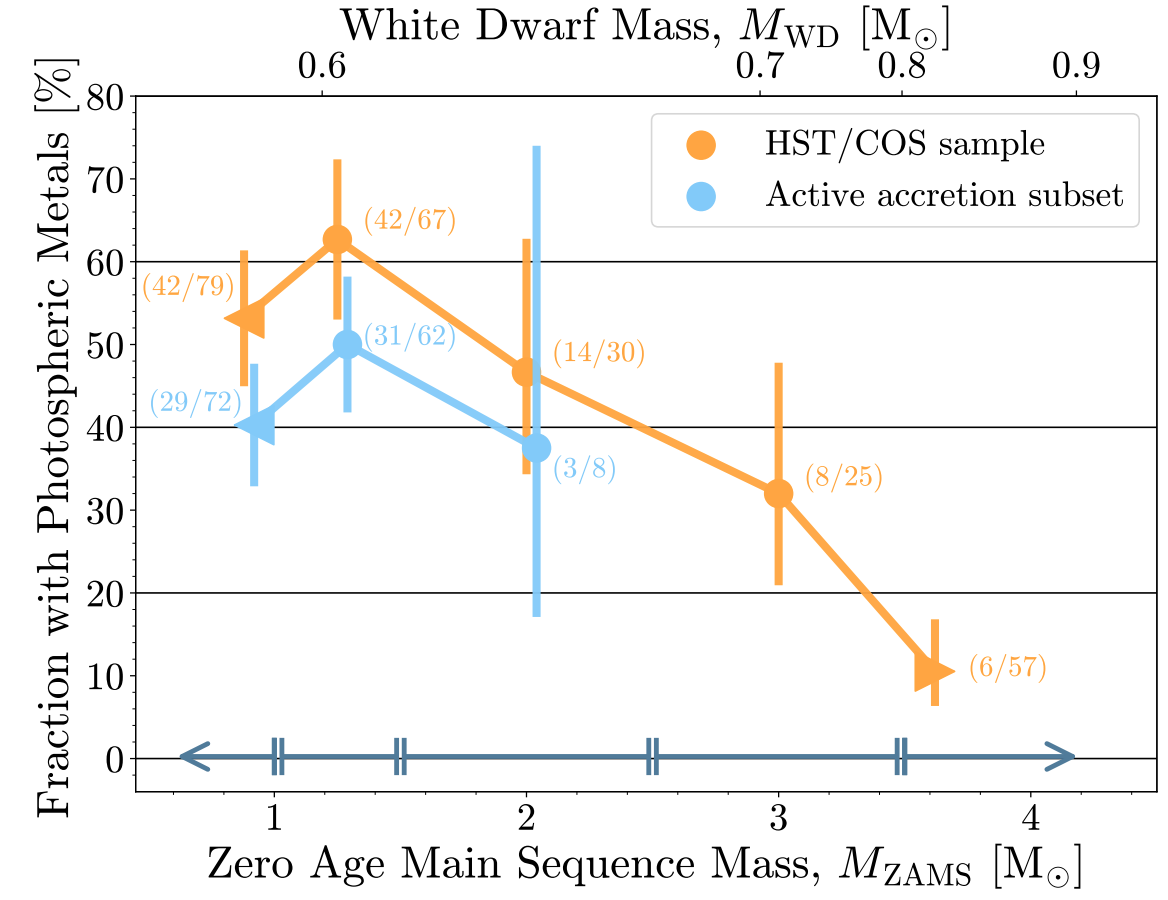}
    \caption{\textit{Left}: The detection and upper limits for Si and Ca as a function of temperature (Sahu et al., in prep). For warm white dwarfs, the Si detection limit is two orders of magnitude lower than the Ca detection limit. Whereas the Ca lines can be detected in the optical, the Si lines can only be detected in the FUV. \textit{Right}: the fraction of white dwarfs with photospheric metals as a function of progenitor mass, showing that fewer massive white dwarfs show evidence of planetary systems \citep[Fig.\,4 from][]{OuldRouis2024}. A larger sample size will increase the robustness of this finding, alongside the capability to do more statistics on planetary system occurrence rates, for example with respect to total system age or binarity.}
    \label{fig:DAZ_detection}
\end{figure}

\vspace{-2mm}
\section*{Volatile detection}
\vspace{-2mm}

The FUV has the strongest transition lines of the volatile elements C, N, O, P, S, with the optical only containing weak oxygen lines. To distinguish dry bodies from wet ones, FUV spectroscopy is required$-$in particular O is crucial because an excess indicates the presence of water ice \citep{Klein2010,Farihi2016,Hoskin2020}. Volatile-rich icy objects from the region analogous to the Kuiper Belt and Oort Cloud are identified by enhancements in C and N \citep{Xu2017,Sahu2025}. The volatile fraction can thus constrain the formation location of planetesimals \citep{Harrison2018}. The C/O ratio of planetesimals has found the frequency of ``carbide planets'' to be $<17\,\%$, albeit only to $2\sigma$ \citep{Wilson2016}, meaning we require a larger sample size. However, in the FUV, volatile elements also have strong interstellar medium (ISM) lines, and thus medium/high resolution spectroscopy is necessary to resolve photospheric and ISM features.

Volatile abundances are also important to constrain exoplanetary core abundances. Although the Earth's core is primarily composed of an Fe--Ni alloy, measurements of the core's density require the presence of a lighter element \citep{Birch1952,Birch1964,Dreibus1996}. Candidate light elements include H, C, O, S, among others. The FUV has strong transitions of siderophiles such as Cr, Fe, and Ni which can measure the core mass fraction of the planetesimal, and several light element candidates. White dwarfs can therefore be used to infer the abundances of light elements in exoplanetary cores \citep{Gaensicke2012,Williams2025}, which is an important input in interior structure models and to infer the interior structure of the Earth.

\vspace{-2mm}
\section*{Long-term ultraviolet monitoring}
\vspace{-2mm}

Many planetary accreting white dwarfs have active, changeable circumstellar environments. Transits from debris change in strength or disappear entirely \citep{Guidry2025} over timescales of days to years \citep{Farihi2022,Aungwerojwit2024}. Gaseous emission lines evolve over months \citep{Wilson2014} and infrared emission also evolves \citep{Swan2019b}. However, the breadth of white dwarfs showing circumstellar variations has not been replicated by white dwarfs with photospheric abundance variability caused by differences in accretion rate. Given that white dwarfs can have sinking timescales as short as $\sim$days--years, we should be sensitive to any changes in accretion rate. Measuring photospheric abundance variability is an empirical test of the modeled sinking timescales \citep[e.g.][]{Koester2009_WDs}. Accretion rate variability will help constrain the evolution of the circumstellar debris disk and would have major implications on the interpretation of accreted debris. To date only one white dwarf has shown evidence of accretion rate variation. Over six epochs spanning 25 years, WD\,0106$-$328 has shown variation in its optical Mg and Ca lines, explained by a reduction in accretion rate of 20\,\% and 60\,\% respectively \citep{Farihi2026}.

Although the variations in WD\,0106$-$328 were detected with optical absorption features, ultraviolet spectroscopy with \textit{HST} is uniquely capable of detecting photospheric abundance variability. Warm ($T_\mathrm{eff}\gtrsim13\,000\,$K) white dwarfs have metal sinking timescales which are on the order of days, compared to millions of years for their cooler counterparts. Therefore, warm white dwarfs are the most sensitive to any accretion rate variation, with the strongest absorption lines in the FUV. In fact, WD\,0106$-$328 is currently undergoing follow-up with \textit{HST} \citep{RogersHST2024}. Long-term ultraviolet spectroscopic monitoring of a sample of warm white dwarfs will determine the frequency of accretion rate variability, which will give us a better understanding of the accretion process. Enriched white dwarfs were first observed with COS in 2010, meaning that if \textit{HST} is operational until 2040, this gives a 30 year baseline, equivalent to thousands of sinking timescales.

Ultraviolet spectroscopy can also detect features from circumstellar gas in emission \citep{LeBourdais2024} and absorption \citep{Dickinson2012}. This can both be used to model the gaseous circumstellar disk \citep{Xu2024} and to search for variation. Upcoming optical spectroscopic surveys will increase the number of gas disk systems by a factor of $\simeq5$ over the next decade \citep{Manser2020}, many of which will require FUV followup. Several white dwarfs show additional \Ion{Si}{IV} circumstellar components \citep{Rogers2024a,Zuckerman2026} which would require a large layer of hot gas to transit the white dwarf. The FUV contains resonant lines for other elements and is thus crucial to understand the gaseous circumstellar environments of planetary accreting white dwarfs.

\section*{Instrument requirements}

To continue the science described in this white paper that is enabled by FUV spectroscopy, we require continued use of COS and STIS. In previous white dwarf studies, the most commonly used COS setting is the G130M grating with central wavelengths 1222\,\AA\ and 1291\,\AA\ because this range contains many strong lines of multiple metals. The COS gratings G160M and G185M also cover strong metal lines. The \textsc{time-tag} mode is very useful in order to probe for short-term variability. Only STIS can obtain an ultraviolet spectrum with a broad wavelength range of $1150-3100\,$\AA, and is vital for high-resolution spectroscopy to resolve components from different origins such as photospheric, interstellar, and circumstellar lines. For the brightest objects in the FUV, only STIS can be used, but the sensitivity of COS is required for fainter targets.

\section*{References}

\bibliographystyle{mnras}
\bibliography{aamnem99,bibliography}


\end{document}